\documentclass[12pt]{article}
\usepackage[OT2,T1]{fontenc}
\usepackage{amsmath,amssymb,epsfig,hyperref}

\makeatletter
\newcommand{\fmslash}[2][0mu]{%
  \mathchoice
    {\fmsl@sh\displaystyle{#1}{#2}}%
    {\fmsl@sh\textstyle{#1}{#2}}%
    {\fmsl@sh\scriptstyle{#1}{#2}}%
    {\fmsl@sh\scriptscriptstyle{#1}{#2}}}
\newcommand{\fmsl@sh}[3]{%
  \m@th\ooalign{$\hfil#1\mkern#2/\hfil$\crcr$#1#3$}}

\title{
\bf A bound on the scale of spacetime noncommutativity from the
reheating phase after inflation
}

\author{
R.Horvat$^1$ and J.Trampetic$^{1,2}$\\
1. Institute Ru\dj er Bo\v skovi\' c,
Bijeni\v{c}ka 54, 10000 Zagreb, Croatia\\
2. Max-Planck-Institut f\"ur Physik, (Werner-Heisenberg-Institut),
  	 \\F\"ohringer Ring 6, D-80805 M\"unchen, Germany\\
E-mail: 
horvat@lei3.irb.hr,
josipt@rex.irb.hr
}

\begin{document}

\maketitle

\begin{abstract}
In an approach to noncommutative gauge theories, where the full
noncommutative behavior is delimited by the presence of the UV and IR
cutoffs, we consider the possibility of describing a system at a temperature
$T$ in a box of size $L$. Employing a specific form of UV/IR relationship
inherent in such an
approach of restrictive noncommutativity, we derive, for a given
temperature $T$, an upper bound on the parameter of spacetime
noncommutativity $\Lambda_{\rm  NC} \sim |\theta|^{-1/2}$. Considering such
epochs in the very early universe which are expected to reflect spacetime
noncommutativity to a quite degree, like the reheating
stage after inflation, or believable pre-inflation radiation-dominated
epochs, the best limits on $\Lambda_{\rm  NC}$ are obtained. We also
demonstrate how the nature and size of the thermal system (for instance,
the Hubble distance versus the future event horizon) can affect our bounds.
\end{abstract}

At the perturbative level noncommutative (NC) quantum field theories (QFT)
suffer from the infamous
UV/IR mixing problem \cite{Minwalla:1999px} and therefore lack
universality \cite{Minwalla:1999px, hepth/0011218, hep-th/0102045}.
This means that any modification at very high
momentum scales does inevitably modify the physics at very small momentum
scales in a profound way, rather than involving a soft modification being
switched off
in the far IR. A treatment of this problem  thus necessitate
understanding on the UV completion of the theory. The formal reason for such
a behavior
is a tacit assumption that in four dimensions NC gauge field theories
are valid up to arbitrarily large momentum scales.

In two important papers \cite{Abel:2006wj,hep-th/0606248},
it was shown that NC gauge theories can
be realized as an effective QFT, underlain by some more fundamental
theory such as string theory. In particular,
it was shown \cite{Abel:2006wj} that at
energy scales below the IR cutoff $\Lambda_{\rm  IR}$,
the NC theory  becomes (up to residual
effects) an ordinary commutative QFT, thereby diminishing substantially the
power of the UV/IR mixing.
On the other hand, it was also claimed \cite{Abel:2006wj} that the
phenomenological effects of the UV completion (for a large class of more
general QFTs above the UV cutoff $\Lambda_{\rm UV}$) can be quite successfully
modelled by a threshold value $\Lambda_{\rm  UV}$. To a good approximation
the theory thus becomes an effective QFT with the UV and IR cutoffs obeying
a relationship
\begin{equation}
\Lambda_{\rm UV} \Lambda_{\rm IR} \sim \Lambda_{\rm NC}^2 \;,
\end{equation}
where the scale of noncommutativity is heuristically introduced as
$\Lambda_{\rm  NC}^{-2} \sim |\theta|$. The full scope of noncommutativity
is experienced only in the range delimited by $\Lambda_{\rm  IR}$ and
$\Lambda_{\rm  UV}$, $\Lambda_{\rm  IR} < \Lambda_{\rm NC} <
\Lambda_{\rm UV}$, while the commutative world (up to residual effects of
noncommutativity) resides below $\Lambda_{\rm  IR}$. Such a scenario had
already been confronted with experimental data
\cite{Abel:2006wj, Horvat:2010km}, and in so doing useful
information on the scale of noncommutativity $\Lambda_{\rm NC}$ was drawn.
Note that with $\Lambda_{\rm NC}$ high enough, the whole standard model can
be placed below $\Lambda_{\rm  IR}$. In this way, one successfully
gets rid of the
Lorentz symmetry violating mass term for photons
\cite{Abel:2006wj}, a relic of the theory in
which the scope of noncommutativity extends up to $\Lambda_{\rm  UV}
\rightarrow \infty$ and down to $\Lambda_{\rm  IR} \rightarrow 0$.

Let us first examine how the above scenario affects the UV/IR mixing
problem. The phenomenon of UV/IR
mixing  is best understood by examining the behavior
of the (nonplanar) loop graphs with the ordinary product of fields replaced
by the Moyal $\star$-product (see e.g.,\cite{hep-th/0106048, hep-th/0109162}).
This results in
phase factors \cite{Filk:1996dm, hep-th/9910004} depending on the
virtual momenta of internal loops. In a theory without UV completion
($\Lambda_{\rm UV} \rightarrow \infty$), these phase factors although
efficient in
damping out the high-energy part of the graphs become together
inefficient to control the vanishing momenta, i.e., the original UV
divergences reappear as IR divergences. On the other hand,
in presence of a finite $\Lambda_{\rm UV}$ no one sort of divergence
will appear since the said phase factors effectively transform the
highest energy
scale ($\Lambda_{\rm UV}$) into the lowest one ($\Lambda_{\rm IR}$). Besides
the appearance of new infrared divergences in the IR limit of the external
momentum, another two (inseparable) aspects \footnote{For a NC field theory
model in which different aspects of the UV/IR mixing problem become
disentangled, see \cite{arXiv:1109.2485}, with details given in
\cite{arXiv:1111.4951}.}  of the UV/IR mixing problem also involve: (i)
nonanalytic behavior in the NC parameter when $\theta \rightarrow 0$ (ii)
pathological behavior when the spatial extension of size $|\theta P|$, for a
particle moving with momentum $P$ along the region affected by spacetime
noncommutativity, gets reduced to a point $|\theta P| \rightarrow 0$. Note
that all aspects of the UV/IR mixing problem get disappeared if NC gauge
theory can be realized as an effective QFT.

The easiest way to understand the peculiar mixing of UV and IR effects in
(1) is to invoke (ii) above, i.e., an interpretation that a quantum in NC
gauge theory can be thought of as a straight string connecting two opposite
charges \cite{Matusis:2000jf, hep-th/9908056, hep-th/9901080}.
Indeed, combining an uncertainty relation for the
coordinates
\begin{equation}
\Delta x^{\mu} \Delta x^{\nu} > \theta^{\mu \nu}\;,
\end{equation}
where $\theta^{\mu \nu}$ has dimensions of $\rm  (length)^2$, with the
Heisenberg uncertainty principle \footnote{Since we are not interested in
the black-hole regime, we do not invoke the Generalized Uncertainty
Principle (GUP) \cite{CERN-TH-5207-88} here.
Yet the holographic principle is discussed below in a
different manner.}, $\Delta x^i \Delta p^j \geq
(1/2)\delta^{ij}$, and switching to the language of effective QFT with
$(\Delta p)_{max} \sim \Lambda_{\rm UV}$, $(\Delta x)_{max} \sim
\Lambda_{\rm IR}^{-1}$ (also employing $\Lambda_{\rm {NC}}^{-2} \sim
|\theta|$), (1) immediately comes about. If we choose  (without loss of
generality) $\theta$ to lie in the (1, 2) plane,
$\theta^{1, 2} = -\theta^{2, 1} \equiv \theta$, this means that a particle
moving inside the NC plane with momentum $P$ along the one axis, has a
spatial extension of size $|\theta P|$ along the other. The string vector is always
perpendicular to the direction of motion.

In the present paper, we require that the effective field-theoretical
treatment of NC gauge theories as given by (1) be also capable of describing a
system at the temperature $T$ in a box of size $L$. We note that an ordinary
effective QFT is expected to be capable of describing a thermal system
provided $\Lambda_{\rm UV} \stackrel{>}\sim  T$ as long as $T \gg L^{-1}$. We shall
focus mainly on radiation-dominated epochs in the very early universe, both
post- and pre-inflationary ones, where the temperature is expected to be
so high that the field-theoretical treatment of such epochs also requires
spacetime noncommutativity. In other words, we require that the above
effective description of spacetime noncommutativity be also capable of describing
radiation-dominated epochs in the early universe. Such a requirement will
bring us a valuable information on the parameter of spacetime
noncommutativity $\Lambda_{\rm {NC}}$.

Sticking to a stringy picture for quanta propagating in NC background, one
immediately sees that having them in thermal equilibrium in the volume $L^3$ is precluded
if the size of the string exceeds the size of the box $L$. The maximal size
of the string in the field-theoretical treatment, for $\theta$ lying in the
(1, 2) plane and averaging over directions of the momentum of the quantum,
can be found to be
\begin{equation}
|\theta p|_{max} = \frac{1}{\sqrt{2}} \frac{\Lambda_{\rm UV}}{\Lambda_{\rm
{NC}}^{2}}\;,
\end{equation}
where $p$ is the total momentum. Thus, the field-theoretical treatment of
spacetime noncommutativity as represented by (1) is expected to be capable
of describing a thermal system (of size $L$) if
\begin{equation}
|\theta p|_{max} \lesssim L \;
\end{equation}
together with $\Lambda_{\rm UV} \stackrel{>}\sim T$ and $T \gg L^{-1}$. This entails
the following upper bound on $\Lambda_{\rm {NC}}$
\begin{equation}
\Lambda_{\rm {NC}} \stackrel{>}\sim (2)^{-1/4} L^{-1/2} T^{1/2}\;.
\end{equation}

In the early universe, the reheating stage marks an abrupt transition from a
cold, low-entropy phase of the inflatory era and a subsequent
high-entropy radiation-dominated epoch
\cite{299778, arXiv:1001.0993}.
In the simplest picture which does not include the preheating
stage after inflation \cite{arXiv:1001.0993, hep-th/9405187, hep-ph/9704452},
the vacuum energy of the inflaton field
experienced an instantaneous conversion into radiation when the decay rate
of the inflaton field had become equal to the expansion rate of the universe
$H$. This event defines the reheating temperature as the maximum temperature of the
subsequent radiation-dominated epoch (although it is not necessarily the
maximum temperature of
the universe after inflation \cite{299778}). At the time
of reheating the Hubble parameter and the reheating  temperature are thus
related as
\begin{equation}
H(T_{RH}) = \left (\frac{8 \pi^3}{90}   g_{*}(T_{RH}) \right)^{1/2}
\frac{T_{RH}^2}{M_{Pl}}\;,
\end{equation}
where $g_{*}$ is the effective number of relativistic degrees of freedom and
$M_{Pl}$ is the Planck mass. The reheating temperature in (6) depends,
through the inflaton decay width, both on
the inflaton mass and on its coupling to matter \cite{299778, arXiv:1001.0993}.

With the most natural choice $L^{-1} = H$ and
using (6), the bound (5) becomes
\begin{equation}
\Lambda_{\rm NC} \stackrel{>}\sim  \left( \frac{4 \pi^{3}}{90}g_{*}(T_{RH}) \right)^{1/4}
\frac{T_{RH}^{3/2}}{M_{Pl}^{1/2}}\;.
\end{equation}
The main reason of why the reheating temperature should not be too high
(thus weakening our bounds) is that one inevitably overproduces gravitinos
in supergravity theories. The limit from gravitino overproduction is $T_{RH}
\lesssim 10^9 - 10^{10}$ GeV \cite{astro-ph/0211258, hep-ph/0307241}.
Taking the effective number of degrees
of freedom at the reheating temperature as for the MSSM ($g_{*}(T_{RH}) = 915/4$)
one obtains for the maximum allowable $T_{RH}$
\begin{equation}
\Lambda_{\rm NC} \stackrel{>}\sim 500 ~ {\rm TeV}\;.
\end{equation}
The bound (8) proves to be as powerful as the bound obtained recently from
nonobservation of ultrahigh energy neutrino induced events in neutrino
observatories \cite{Horvat:2010sr}.

Next we demonstrate how the heuristic arguments leading to (7) can be beefed
up by invoking entropic considerations, and, in particular, the holographic
bound. For a collection of strings of length $\sim$ $\Lambda_{\rm {IR}}^{-1}$
in the volume $\sim$ $\Lambda_{\rm {IR}}^{-3}$, the entropy is bounded by the
Bekenstein bound $S_B$ \cite{NSF-ITP-80-38}. For a macroscopic system in which
self-gravitation effects can be disregarded, the Bekenstein bound is given
by a product of the energy and the linear size of the system, $EL$. In the
context of effective QFTs, it becomes proportional
$\Lambda_{\rm {UV}}^4 \Lambda_{\rm {IR}}^{-4}$. It should be noted that it
is more extensive than the entropy in effective QFTs,
$S_{QFT} \sim \Lambda_{\rm {UV}}^3
\Lambda_{\rm {IR}}^{-3}$. On the other hand,
for a weakly gravitating system $S_B$ is bounded
by the holographic Belkenstein-Hawking entropy, $S_{BH} \sim \Lambda_{\rm
{IR}}^{-2}M_{Pl}^2$ \footnote{It was shown \cite{Horvat:2010km} that in the regime
obeyed by the present field-theoretical framework, NC thermodynamical laws
are a NC deformation of the usual laws \cite{arXiv:0801.3583, gr-qc/0510112}.
Thus the commutative area law,
$S_{BH}^{\rm NC} = A M_{Pl}^2/4$, stays preserved in a NC setting.}.
Ignoring, for simplicity, the numerical factors,
setting \footnote{Setting, on the other hand, $S_{QFT} \leq S_B$ and
invoking (1) gives us a consistency condition for the theory, $\Lambda_{\rm
{UV}} \stackrel{>}\sim \Lambda_{\rm {NC}} \stackrel{>}\sim \Lambda_{\rm {IR}}$.
The final option,
$S_{QFT} \leq S_{BH}$, means that with this bound, at saturation, our
effective theory should also be
capable of describing systems containing black holes, since it necessarily
includes many states with Schwarzschild radius much larger than the box
size. There are however compelling arguments for why an ordinary local effective
QFT appears unlikely to provide an adequate description of any systems
containing black holes.}
\begin{equation}
S_B \leq S_{BH} \;,
\end{equation}
and invoking (1), one arrives at
\begin{equation}
\Lambda_{\rm {NC}} \stackrel{>}\sim \frac{\Lambda_{\rm {UV}}^{3/2}}{M_{Pl}^{1/2}}\;.
\end{equation}
Such a framework is capable of describing a thermal system if $\Lambda_{\rm
{UV}} \stackrel{>}\sim T$, where $T \gg L^{-1}$. This way, (7) is immediately reproduced.
Having employed $\Lambda_{\rm {UV}} \stackrel{>}\sim \Lambda_{\rm {NC}}$ in (10), we
obtain that only subplanckian
noncommutativity, $\Lambda_{\rm {NC}} \lesssim M_{Pl}$, is allowed by the
holographic bound. In
fact, entropic considerations employed here generalize the bound
(5) in such a way that $L$ is being replaced with the largest possible size
of the string consistent
with the holographic bound at the temperature $T$. In radiation-dominated
cosmologies, such a scale is provided by the
Hubble distance, although other choices for $L$ are also possible (see below). Now we are
on the much firmer ground with our bounds.

One may object, though, that our bounds include spacetime  noncommutativity on the
particle theory side only, and do not consider the possibility that
noncommutativity can affect particular epochs in the history of the universe or even the whole history as well
\cite{hep-th/0408071, hep-th/0407111,  arXiv:1003.1194, arXiv:1011.3896,
arXiv:1109.3514}.
Since our interest is in
radiation dominated epochs in the early universe, the only real concern is a fact that
NC spacetime geometry leads to modified dispersion relations \cite{hep-th/0108190},
affecting in turn various thermodynamical quantities. These become dependent
not only on the temperature, but also on the parameter characterizing
spacetime noncommutativity. Even so, our effective
treatment is expected to provide an adequate description of such systems as long
as the hierarchy $\Lambda_{\rm {UV}} \stackrel{>}\sim \Lambda_{\rm {NC}}, T$ is
respected. In particular, a modification of  dispersion relations  for the radiation
has already been studied \cite{Abel:2006wj} in the  effective field-theoretical model
obeying (1). While at low momentum scales $(k \ll \Lambda_{\rm {IR}})$ one gets a
polarization dependent propagation speed (birefringence), for $\Lambda_{\rm
{IR}} \ll k \ll \Delta M_{\rm {SUSY}}$ a Lorentz violating mass term $\sim
\Delta M_{\rm {SUSY}}$ emerges, where $\Delta M_{\rm {SUSY}}$ is the supertrace of the
mass matrix. Thus at high temperatures of interest here, $T \gg \Delta M_{\rm {SUSY}}$, a
modification to ordinary dispersion relations turns out to be negligible.

A possibility of having a radiation-dominated epoch taken place before inflation is
quite generic in many extensions of the $\Lambda$CDM model, especially in those
with a symmetry-breaking phase transition characterized by a high-energy
scale (e.g., GUT symmetry breaking) \cite{299778}. The solution to such a
transition has been obtained long ago \cite{TUTP-82-7}. Recently, the effect of
such a pre-inflation radiation-dominated epoch to CMB anisotropy has been
studied \cite{astro-ph/0612006, arXiv:0704.2095}. Even the inflation itself may
occur in a state with thermal relativistic matter, when the equation of state
turns into that of inflationary matter under influence of the NC structure
of spacetime \cite{hep-th/0108190}. All such thermal epochs are expected to be
describable
altogether by our NC effective QFT approach if the hierarchy $\Lambda_{\rm
{UV}} \stackrel{>}\sim \Lambda_{\rm {NC}}, T$ is respected, supplemented also by a
holographic bound $\Lambda_{\rm {UV}} \lesssim M_{Pl}$. Having the temperature
in these epochs
much higher than in the phase of reheating, the bound on
the scale of noncommutativity (8) is upraised by a tremendous amount.
For instance, for a GUT scale of about $10^{15}$ GeV one would have
$\Lambda_{\rm {NC}} \stackrel{>}\sim 10^{10}$ TeV. For a radiation-dominated epoch near
the Planck epoch, the bound is obviously saturated by the holographic
constraint near the Planck mass. It is important to note that our
bounds do not require any prior specification of the cutoffs as long as the aforementioned
hierarchy between the scales stays respected.

Finally, we would like to see to what extent choices different than
$L^{-1} = H$ could affect our bound (8). For that purpose we resort to a
particular cosmological model, which, as a bonus, is also proven to be successful in
describing
the present accelerated phase of the universe. We choose the model for
holographic dark energy
\cite{hep-th/0403052, Li:2004rb, astro-ph/0404204}, a stuff prevailing at present times but
suppressed at earlier cosmological epochs. For the sake of
illustration, we consider the popular Li's model \cite{Li:2004rb}. This model
belongs to a class of
noninteracting and saturated HDE models, with a choice for $L$ in the form
of the future event horizon,
\begin{equation}
d_{E} = a \int_{a}^{\infty } \frac{da}{a^2 H} \;,
\end{equation}
with $a$ being a scale factor. The vacuum energy density, assumed not to be
responsible for the early-time inflation,  is parametrized as
$\rho_{\Lambda } =(3/8 \pi ) M_{Pl}^2 d_{E}^{-2}$. Extracting $d_{E}$
amounts to knowing $\rho_{\Lambda }$ during the radiation-domination epoch,
in which $\rho_{\Lambda }$ occupies only a tiny fraction of the total energy
density. In a
two-component universe $\rho_{\Lambda}$ evolution is governed
by \cite{Li:2004rb, hep-th/0506069} \footnote{The modification of the right hand side of the
Einstein equations arising from the fuzziness of space induced by
$\Lambda_{\rm {NC}}$ \cite{arXiv:0908.1949} can be shown to be unimportant for the
radioation-dominated  epochs of interest here.}
\begin{equation}
\Omega_{\Lambda }^{'} = \Omega_{\Lambda }^2 ( 1 - \Omega_{\Lambda }) \left
[\frac{1}{\Omega_{\Lambda }} + \frac{2}{\sqrt{\Omega_{\Lambda }}} \right ]\;,
\end{equation}
where the prime denotes the derivative with respect to $\rm {lna}$. In (12)
$\Omega_{\Lambda } = \rho_{\Lambda }/\rho_{crit} $, where $\rho_{crit}$ is
the critical density. With $ \Omega_{\Lambda } << 1$ and $\rho_{crit} \simeq
\rho_{rad }$, we obtain
\begin{equation}
\rho_{\Lambda }(a) \simeq g_{*}(a) ~\rho_{rad_0} ~ a^{-3} \;,
\end{equation}
where $\rho_{rad_0}$  denotes the radiation energy density at the present
time. In turn this, together with a solution of (12) for the matter-dominated epoch,
\begin{equation}
\rho_{\Lambda }(a) \simeq \rho_{m_0}~ a^{-2} \;,
\end{equation}
determines the ratio $L^{-1/2}/H^{1/2}$ at the time of reheating as
\begin{equation}
\frac{L^{-1/2}(T_{RH})}{H^{1/2}(T_{RH})} \sim \frac{T_{0}}{T_{RH}^{1/4}
T_{MR}^{3/4}}  \;,
\end{equation}
where $T_{0}$ is the temperature of the universe at present and $T_{RM}$ is
the temperature at the moment when the matter density becomes equal to that
of the radiation. Plugging some relevant numbers and $T_{RH}=10^{10}$ GeV,
one finds the ratio (14) to be $\sim 10^{-10}$. This leads, via (5), to a
insignificant bound on $\Lambda_{\rm {NC}}$. As has been already made clear with (7),
the holographic bound on $\Lambda_{\rm {NC}}$ virtually coincides with the bound (5)
for the choice $L = H^{-1}$. Thus any choice for $L$ being larger  than the
Hubble distance at the time of reheating leads to a weaker bound. This arguably demonstrates
how strongly the choice for the 'size' of the universe
can influence our bounds. There are of course other choices for $L$ being
relevant
for the bound (5); for instance, the particle horizon. For this choice for
$L$ one
would expect a bound similar to (8); however, the employed model in this
case  can no longer be responsible for the current accelerated phase of the
universe.

Summing up, we have made use of a reasonable expectation that thermal epochs
in the universe are successfully describable by conventional QFTs. We have
considered the field-theoretical realization of noncommutative gauge field
theories and shown, making use of the specific UV/IR relationship attributive to such
an approach, that adequate description of a thermal system entails an upper
bound on the scale of noncommutativity. We have also discussed such a bound
in conjuction with the holographic bound. For the reheating stage after
inflation we have obtained a bound of order of $10^{3}$ TeV. The
radiation-dominated epochs at higher temperatures, if existent, would
provide much better bounds, possibly all the way up to the Planck mass.
It is important to notice that only existence of such epochs matters and not
the physics responsible for their emergence, since in thermal equilibrium the
physical system loses  memory of its initial state. We
have also demonstrated how the size and nature of the thermal system may
crucially affect our bounds.

J.T. would like to acknowledge support of
Alexander von Humboldt Foundation
(KRO 1028995), W. Hollik and Max-Planck-Institute for Physics, Munich, for hospitality.
The work of R.H. and J.T. are supported by
the Croatian Ministry of Science, Education and Sports
under Contracts Nos. 0098-0982930-2872 and 0098-0982930-2900, respectively.

\end{document}